\documentclass[conference]{IEEEtran}

\usepackage{bm}
\usepackage{caption}
\usepackage{subcaption}
\usepackage{graphicx}
\usepackage{cite}
\usepackage{amsmath,amsfonts}
\begin{document}

%\title{Probabilistic Models for Electronic Health Records and Inference Under Partial Observations with Application to Resource Planning}
%\title{Probabilistic Model and Inference Algorithms for Electronic Health Record Sequences}
\title{Sequential Inference of Hospitalization Electronic Health Records Using Probabilistic Models}

\author{\IEEEauthorblockN{Alan D. Kaplan and Priyadip Ray}
\IEEEauthorblockA{
%\textit{Computational Engineering Division} \\
\textit{Lawrence Livermore National Laboratory}\\
Livermore, USA}
\and
\IEEEauthorblockN{John D. Greene and Vincent X. Liu}
\IEEEauthorblockA{
%\textit{dept. name of organization (of Aff.)} \\
\textit{Kaiser Permanente Division of Research}\\
Oakland, USA}
}

%\lhead{DRAFT 2023-05-02}

\maketitle

\begin{abstract}%   <- trailing '%' for backward compatibility of .sty file
%Inference of future outcomes
%Heterogeneity and irregular sampling make modeling of Electronic Health Record (EHR) data difficult.
In the dynamic hospital setting, decision support can be a valuable tool for improving patient outcomes.
Data-driven inference of future outcomes is challenging in this dynamic setting, where long sequences such as laboratory tests and medications are updated frequently.
This is due in part to heterogeneity of data types and mixed-sequence types contained in variable length sequences.
In this work we design a probabilistic unsupervised model for multiple arbitrary-length sequences contained in hospitalization Electronic Health Record (EHR) data.
The model uses a latent variable structure and captures complex relationships between medications, diagnoses, laboratory tests, neurological assessments, and medications.
It can be trained on original data, without requiring any lossy transformations or time binning.
Inference algorithms are derived that use partial data to infer properties of the complete sequences, including their length and presence of specific values.
We train this model on data from subjects receiving medical care in the Kaiser Permanente Northern California integrated healthcare delivery system.
The results are evaluated against held-out data for predicting the length of sequences and presence of Intensive Care Unit (ICU) in hospitalization bed sequences.
Our method outperforms a baseline approach, showing that in these experiments the trained model captures information in the sequences that is informative of their future values.
%Our method is compared to a population average baseline that show in these experiments our approach captures sequence information that is informative of future values.

\end{abstract}

LLNL-TR-849343

%\thispagestyle{firstpage}

%\fancyhf{} % sets both header and footer to nothing
%\renewcommand{\headrulewidth}{0pt}
% your new footer definitions here

\section{Introduction}
%The collection and analysis of Electronic Health Record (EHR) data is central to the aims of precision medicine \cite{Ginsburg2018-zf, Kosorok2019-bz}.
%Patterns that lie within Electronic Health Record (EHR) data may assist in the formulation of individualized medical decision making \cite{Kim2019-bo}.
%Common characteristics of EHR data that make data-driven analysis challenging include non-uniformly sampled and variable length sequences, such as laboratory tests and medications.

Many different machine learning approaches have been explored in the space of EHR modeling and prediction (see, e.g. \cite{Wu2010-hr, Shickel2018-qy}).
Methods have been developed for use in disease prediction \cite{Chen2017-re, Kohli2018-fj, Uddin2019-xm}, length of stay prediction \cite{Gustafson1968-xh, Awad2017-me}, biomarker discovery \cite{Huynh-Thu2012-fa, Ledesma2021-nl}, treatment planning \cite{Nicolae2017-ix, Valdes2017-jy}, and subgroup analysis \cite{Vranas2017-gn, Pinal-Fernandez2020-kn, Wang2020-ck, Kaplan2022-fo}.
These approaches also vary with the type of input data used, which can include summary statistics, images, time series, and sequences.

For this work, we construct a probabilistic model of sequences contained in episodic EHR data collected during hospitalization.
Probabilistic methods allow us to compute distributions of quantities of interest, which is useful in an uncertain clinical environment.
In addition, one model can be trained over all of the data and used for multiple inference tasks by computing conditional probabilities of interest.
Modeling of multiple sequences, such as medications and laboratory tests, has several applications, including prognosis of patient health trajectories, resource allocation, treatment planning, and outcome prediction.
A single model that can perform multiple tasks is beneficial in a dynamic clinical environment, where both the nature of the known data and desired predictive outcomes may be changing often.

%For this work, we are focused on constructing a probabilistic model of episodic EHR data collected during hospitalization.
%Sources of uncertainty inherent in data-driven decision support include noisy measurements and weak correlations between clinical measures and outcomes.
%Probabilistic methods allow us to compute distributions of quantities of interest rather than point predictions, which is useful in an uncertain clinical environment.
%In addition, one model can be trained over all of the data and used for multiple inference tasks by computing conditional probabilities of interest.
%A single model that can perform multiple tasks is beneficial in a dynamic clinical environment, where both the nature of the known data and desired predictive outcomes may be changing often.
%In general, a variety of tools and approaches may be required to produce the highest quality inference from data-driven approaches.

Our goal for this work is to develop a flexible approach that can predict properties of future values of a patient's EHR given partial sequences of their EHR, using a single probabilistic model.
This requires the construction of the underlying probability model and derivation of inference algorithms that use this model.

We use the formulation of mixture models, which have been previously investigated in the context of EHR data \cite{McLachlan2004-mh, Najjar2015-rb, Cheung2017-gk, Kaplan2022-fo}.
Both probabilistic and neural network based approaches have been used that incorporate sequential dynamics of EHR data \cite{Stella2012-vr, Liu2015-ad, Futoma2017-os, Alaa2017-nd, Kaplan2022-fo, Choi2016-pr, Jin2018-rh}.
In addition, combinations of probabilistic subgrouping methods and dynamic models have been developed \cite{Alaa2018-nq, Cui2022-pn, McDowell2018-ae, Kaplan2021-ym}.
In contrast with this work, these methods do not incorporate sequences with simultaneous observations (such as multiple medications administered at one timepoint).
Also, these prior works do not derive general inference methods that can be used to estimate arbitrary future values of EHR sequences.
Prior work that developed mixture models for EHR data showed that learned subgroup phenotypes can be used to gain insight from sequences in EHR data \cite{Kaplan2022-fo}.
The model in the present work utilizes a mixture model at the top level, but differs in its latent variable structure.
In addition, in this work we focus on inference, by deriving expressions for distributions of future values and perform validation of prediction.

The model developed in this work is defined by a full joint probability distribution over the data components and can fit the data without requiring any lossy transformations.
Inference procedures are derived for estimating final sequence lengths at discharge and future values using conditional likelihoods from the original model.
Once trained, we evaluate the performance of the method on data from Kaiser Permanente Northern California (KPNC).
We perform these tasks at both the individual and population-levels and show that they improve performance beyond baseline population statistic approaches.

\section{Data}
KPNC is a highly integrated healthcare delivery system with 21 medical centers caring for an overall population of 4 million members.
For this work, we use a dataset consisting of 244,248 inpatient hospitalization visits with a suspected or confirmed infection and sepsis diagnosis, drawn from KPNC medical centers between 2009 and 2013 \cite{Liu2017-al}.

We consider the following data elements:
\begin{itemize}
    \item \textsf{Age}: $\phi_a$
    \item \textsf{Sex}: $\phi_s$
    \item \textsf{Death}: $\phi_d$
    \item \textsf{Beds}: $\boldsymbol{\beta} = \{\beta_1, \ldots, \beta_{|\boldsymbol{\beta}|}\}$
    \item \textsf{Admission Diagnoses}: $\boldsymbol{\alpha} = \{\alpha_1, \ldots, \alpha_{|\boldsymbol{\alpha}|} \}$
    \item \textsf{Discharge Diagnoses}: $\boldsymbol{\delta} = \{\delta_1, \ldots, \delta_{|\boldsymbol{\delta}|} \}$
    \item \textsf{Laboratory Tests}: $\boldsymbol{\lambda} = \{\boldsymbol{\lambda}_1, \ldots, \boldsymbol{\lambda}_{|\boldsymbol{\lambda}|}\} = 
    \{
    \{\lambda_{1, 1}, \ldots, \lambda_{1, |\boldsymbol{\lambda}_1|} \},
    \ldots,
    \{\lambda_{|\boldsymbol{\lambda}|, 1}, \ldots, \lambda_{|\boldsymbol{\lambda}|, |\boldsymbol{\lambda}_{|\boldsymbol{\lambda}|}|} \}
    \}$
    \item \textsf{Neurological Tests}: $\boldsymbol{\nu} = \{\boldsymbol{\nu}_1, \ldots, \boldsymbol{\nu}_{|\boldsymbol{\nu}|}\} = 
    \{
    \{\nu_{1, 1}, \ldots, \nu_{1, |\boldsymbol{\nu}_1|} \},
    \ldots,
    \{\nu_{|\boldsymbol{\nu}|, 1}, \ldots, \nu_{|\boldsymbol{\nu}|, |\boldsymbol{\nu}_{|\boldsymbol{\nu}|}|} \}
    \}$
    \item \textsf{Medications}: $\boldsymbol{\mu} = \{\boldsymbol{\mu}_1, \ldots, \boldsymbol{\mu}_{|\boldsymbol{\mu}|}\} = 
    \{
    \{\mu_{1, 1}, \ldots, \mu_{1, |\boldsymbol{\mu}_1|} \},
    \ldots,
    \{\mu_{|\boldsymbol{\mu}|, 1}, \ldots, \mu_{|\boldsymbol{\mu}|, |\boldsymbol{\mu}_{|\boldsymbol{\mu}|}|} \}
    \}$
\end{itemize}

Bold symbols indicate vectors of values.
In cases where we refer to any of these elements, we use the symbol $\boldsymbol{x}$.
The number of values in $\boldsymbol{x}$ is $|\boldsymbol{x}|$.
\textsf{Admission Diagnoses} and \textsf{Discharge Diagnoses} are collections of ICD-9 codes that are given apon admission and discharge, respectively.
Sequential information that are collected throughout the each episode include \textsf{Beds}, \textsf{Laboratory Tests}, \textsf{Neurological Tests}, and \textsf{Medications}.
These sequences are of varying length and may contain multiple observations of varying quantity at each timepoint.
%Note that for \textsf{Laboratory Tests}, \textsf{Neurological Tests}, and \textsf{Medications}, more than one item is possible at each time point.
%This means that $\boldsymbol{\lambda}, \boldsymbol{\nu}$, and $\boldsymbol{\mu}$ may contain sequences of varying length.
For these sequences, the time associated with a value is denoted $t_{x, i}$.
For example, the timestamp (in hours) of the collection of laboratory tests $\boldsymbol{\lambda_i}$ is $t_{\lambda, i}$ and the timestamp for the medications $\boldsymbol{\mu_i}$ is $t_{\mu, i}$.
%The \textsf{Age}, \textsf{Sex}, and \textsf{Death} variables do not have timestamps associated with them.
The collection of data for one episode is: $\boldsymbol{y}=\left(\phi_a, \phi_s, \phi_d, \boldsymbol{\beta}, \boldsymbol{\alpha}, \boldsymbol{\delta}, \boldsymbol{\lambda}, \boldsymbol{\nu}, \boldsymbol{\mu} \right)$.

\section{Methods} \label{sec:methods}

\subsection{Model structure}

\begin{figure*}[ht]
        \includegraphics[width=0.8\textwidth]{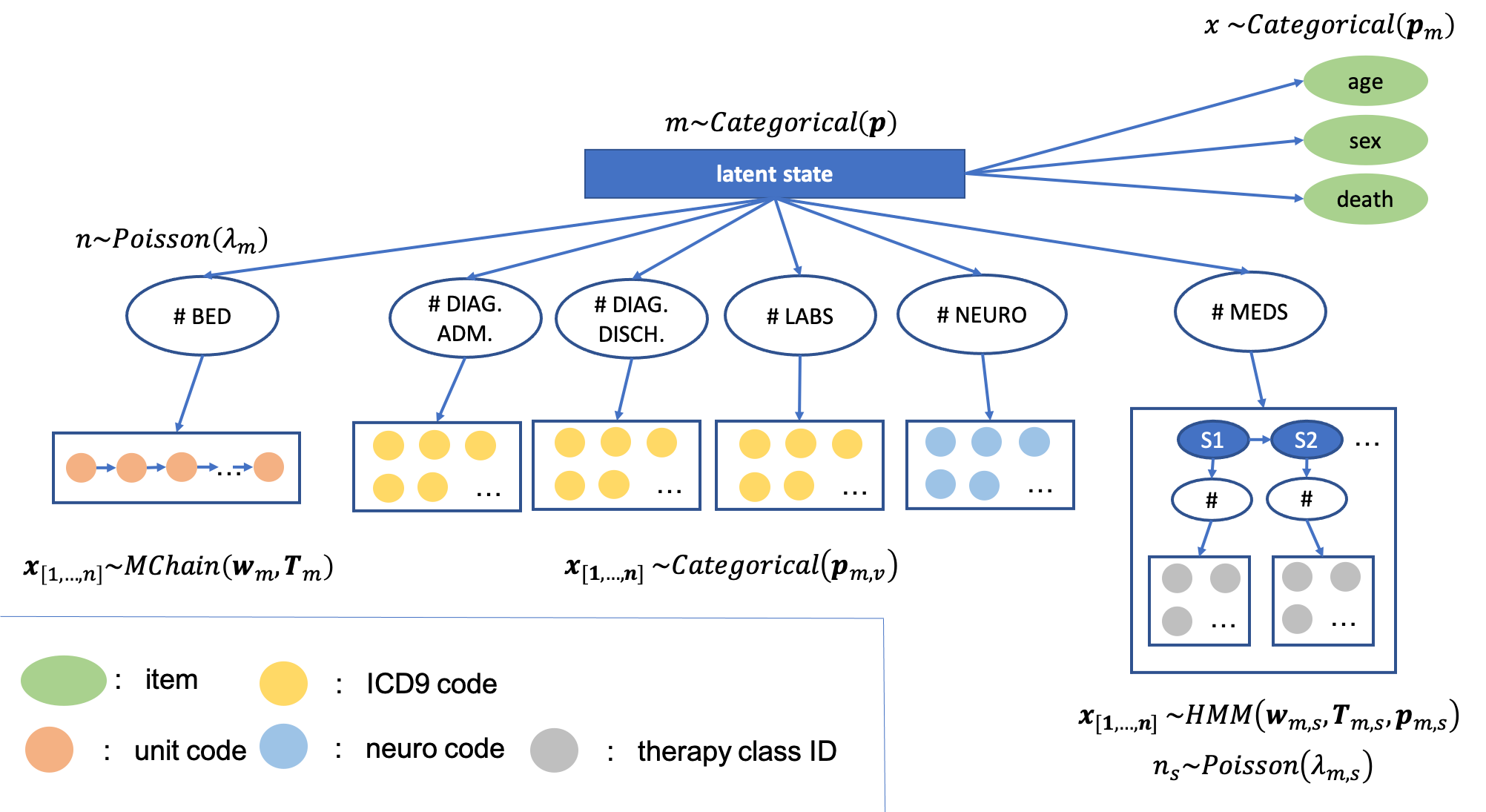}
        \centering
        \caption{This figure shows a graphical representation of the model. Arrows indicate dependencies incorporated the model. Each colored circle represents a single value of data.}
        \label{fig:model}
\end{figure*}

In this section we describe the probabilistic model that computes the likelihood of an episode, $f\left(\boldsymbol{y}\right)$.
%Our model in generative and works with irregularly spaced sequences and their combination with scalar elements.
Figure \ref{fig:model} shows the structure of the model.
Colored circles represents single values from the data elements and arrows indicate probabilistic dependencies.

The model utilizes a mixture model formulation with a latent state $z$ that characterizes associations between data elements.
%, and is drawn from a random variable $Z$.
The data elements of the episode are independent conditioned on the latent variable, so that
    \begin{equation} \label{eq:condprob}
        f\left(\boldsymbol{y} | Z=z \right) = f\left(\phi_a | Z=z\right)f\left(\phi_s | Z=z\right) \cdots f\left(\boldsymbol{\mu} | Z=z\right),
    \end{equation}
and the distribution of the model is formed by summing over the latent variable,
    $$
        f\left(\boldsymbol{y}\right) = \sum_z p_z f\left(\boldsymbol{y} | Z=z \right).
    $$
%Note that since the latent variable is unknown, the components of the episode are not independent under the model.
%For example, the likelihood of \textsf{Age} given \textsf{Sex} under this model is $f\left(\phi_a | \phi_s\right) = \frac{\sum_z p_z f\left(\phi_a | Z = z \right)f\left(\phi_s | Z = z \right)}{\sum_z p_z f\left(\phi_s | Z = z \right)} \propto \sum_z p_z f\left(\phi_a | Z = z \right)f\left(Z = z | \phi_s \right) \neq f\left(\phi_a\right)$.
%If the model treated all components to be independent, then this would evaluate to the marginal, $f\left(\phi_a\right)$, without considering \textsf{Sex}.

Each product term in Eq \ref{eq:condprob} is specified with distribution families that match the corresponding data types.
%Now we specify the conditional distributions for each of the product terms.
For \textsf{Age}, we use a quantized and truncated Gaussian distribution,
%While the Poisson may be more natural, we find that variance increasing with the mean is too restrictive.
%This conditional distribution is,
    $$
        f\left(\phi_a | Z=z\right) = \frac{1}{C\sqrt{2\pi\sigma_z^2}}e^{\frac{\left(m_z - \phi_a\right)^2}{2\sigma_z^2}},
    $$
where the normalizing value ensures that the distribution sums to 1, $C=\sum_{\phi_a}\frac{1}{\sqrt{2\pi\sigma_z^2}}e^{\frac{\left(m_z - \phi_a\right)^2}{2\sigma_z^2}}$.

For \textsf{Sex} and \textsf{Death}, we we use Bernoulli distributions,
    $$
        f\left(\phi_s | Z=z\right) = p_{s,z}^{1 - \phi_s}
        \left(1 - p_{s,z}\right)^{\phi_s},
    $$
    $$
        f\left(\phi_d | Z=z\right) = p_{d,z}^{1 - \phi_d}
        \left(1 - p_{d,z}\right)^{\phi_d}.
    $$
    
For \textsf{Admission Diagnoses}, \textsf{Discharge Diagnoses}, \textsf{Laboratory Tests}, and \textsf{Neurological Tests}, we use products of categorical (i.e. sometimes referred to as ``bag-of-words") models.
The length of each of these sequences for the entire episode (e.g. $|\boldsymbol{\alpha}|$ for \textsf{Admission Diagnoses} and $\sum_{i=1}^{|\boldsymbol{\lambda}|} |\boldsymbol{\lambda}_i|$ for \textsf{Laboratory Tests}) is modeled using a Poisson distribution.
For these sequences we have,
    %$$
    %    f\left(\boldsymbol{\alpha} | Z=z\right) = 
    %    \frac{l_{\alpha, z}^{|\boldsymbol{\alpha}|}}{|\boldsymbol{\alpha}|!}e^{-l_{\alpha, z}}
    %    p_{\alpha,1}^{N_{\alpha,1}} p_{\alpha,2}^{N_{\alpha,2}} \cdots p_{\alpha,|\mathcal{C_\alpha}|}^{N_{\alpha,|\mathcal{C_\alpha}|}},
    %$$
    $$
        f\left(\boldsymbol{\alpha} | Z=z\right) = 
        \frac{l_{\alpha, z}^{|\boldsymbol{\alpha}|}}{|\boldsymbol{\alpha}|!}e^{-l_{\alpha, z}}
        \prod_{i=1}^{|\boldsymbol{\alpha}|} p_{\alpha, z, \alpha_i},
    $$
    %$$
    %    f\left(\boldsymbol{\delta} | Z=z\right) = 
    %    \frac{l_{\delta, z}^{|\boldsymbol{\delta}|}}{|\boldsymbol{\delta}|!}e^{-l_{\delta, z}}
    %    p_{\delta,1}^{N_{\delta,1}} p_{\delta,2}^{N_{\delta,2}} \cdots p_{\delta,|\mathcal{C_\delta}|}^{N_{\delta,|\mathcal{C_\delta}|}},
    %$$
    $$
        f\left(\boldsymbol{\delta} | Z=z\right) = 
        \frac{l_{\delta, z}^{|\boldsymbol{\delta}|}}{|\boldsymbol{\delta}|!}e^{-l_{\delta, z}}
        \prod_{i=1}^{|\boldsymbol{\delta}|} p_{\delta, z, \delta_i},
    $$
    %$$
    %    f\left(\boldsymbol{\lambda} | Z=z\right) = 
    %    \frac{l_{\lambda, z}^{\sum_{i=1}^{|\boldsymbol{\lambda}|} |\boldsymbol{\lambda}_i|}}{\sum_{i=1}^{|\boldsymbol{\lambda}|} |\boldsymbol{\lambda}_i|!}e^{-l_{\lambda, z}}
    %    p_{\lambda,1}^{N_{\lambda,1}} p_{\lambda,2}^{N_{\lambda,2}} \cdots p_{\lambda,|\mathcal{C_\lambda}|}^{N_{\lambda,|\mathcal{C_\lambda}|}},
    %$$
    $$
        f\left(\boldsymbol{\lambda} | Z=z\right) = 
        \frac{l_{\lambda, z}^{\sum_{i=1}^{|\boldsymbol{\lambda}|} |\boldsymbol{\lambda}_i|}}{\sum_{i=1}^{|\boldsymbol{\lambda}|} |\boldsymbol{\lambda}_i|!}e^{-l_{\lambda, z}}
        \prod_{i=1}^{|\boldsymbol{\lambda}|} 
        \prod_{j=1}^{|\boldsymbol{\lambda}_i|}
        p_{\lambda, z, \lambda_{i,j}},
    $$
    %$$
    %    f\left(\boldsymbol{\nu} | Z=z\right) = 
    %    \frac{l_{\nu, z}^{\sum_{i=1}^{|\boldsymbol{\nu}|} |\boldsymbol{\nu}_i|}}{\sum_{i=1}^{|\boldsymbol{\nu}|} |\boldsymbol{\nu}_i|!}e^{-l_{\nu, z}}
    %    p_{\nu,1}^{N_{\nu,1}} p_{\nu,2}^{N_{\nu,2}} \cdots p_{\nu,|\mathcal{C_\nu}|}^{N_{\nu,|\mathcal{C_\nu}|}},
    %$$
    $$
        f\left(\boldsymbol{\nu} | Z=z\right) = 
        \frac{l_{\nu, z}^{\sum_{i=1}^{|\boldsymbol{\nu}|} |\boldsymbol{\nu}_i|}}{\sum_{i=1}^{|\boldsymbol{\nu}|} |\boldsymbol{\nu}_i|!}e^{-l_{\nu, z}}
        \prod_{i=1}^{|\boldsymbol{\nu}|} 
        \prod_{j=1}^{|\boldsymbol{\nu}_i|}
        p_{\nu, z, \nu_{i,j}},
    $$
%where $N_{\cdot, i}$ is the number of occurrences of item $i$ in the $\cdot$ sequence and $|\mathcal{C_\cdot}|$ is the number of possible items in the $\cdot$ sequence.
where $p_{x, z, i}$ is the probability of item $i$ in sequence $x$ for latent state $z$.

For the \textsf{Beds} sequence we use a Markov Chain to model transitions between timepoints, so that the order of observations is significant.
A Poisson distribution characters the length of the sequence, leading to,
    $$
        f\left(\boldsymbol{\beta} | Z=z\right) = 
        \frac{l_{\beta, z}^{|\boldsymbol{\beta}|}}{|\boldsymbol{\beta}|!}e^{-l_{\beta, z}}
        p_{\beta,z, \beta_1} \prod_{i=1}^{|\boldsymbol{\beta}| - 1} q_{\beta, z, \beta_i, \beta_{i+1}},
    $$
where $p_{\beta,z,i}$ is the probability that the first item is $i$ conditioned on $Z=z$, and $q_{\beta, z, i, j}$ is the probability of transitioning from item $i$ to $j$ conditioned on $Z=z$.

We also capture the order and transitions between timepoints in the \textsf{Medications} sequence.
However, unlike the \textsf{Beds} sequence, the \textsf{Medications} can contain multiple observations at any single timepoint.
%A Markov Chain is not sufficient to characterize this feature of the data.
%In order to allow for this feature, we utilize an extension to the Markov Chain model.
We use a sequence of latent states, $\boldsymbol{S}$, where each of these states carries a conditional distribution over a collection of medications of arbitrary length.
In this way we are able to model the ordered sequence and express multiple simultaneous observations using a Hidden Markov Model with variable length observations for each state, $s$, characterized by Poisson distributions.
The distribution is,
    \begin{multline*}
        f\left(\boldsymbol{\mu} | Z=z\right) = 
        \frac{l_{\mu, z}^{|\boldsymbol{\mu}|}}{|\boldsymbol{\mu}|!}e^{-l_{\mu, z}}
        \sum_{\boldsymbol{S}}
        \left(
        p_{\mu,z,s_1} \prod_{i=1}^{|\boldsymbol{\mu}| - 1} q_{\mu,z, s_i, s_{i+1}}
        \right) \\
        \left(
        \prod_{i=1}^{|\boldsymbol{\mu}|}
        \frac{l_{\mu, z, s_i}^{|\boldsymbol{\mu}_i|}}{|\boldsymbol{\mu}_i|!}e^{-l_{\mu, z, s_i}}
        \prod_{j=1}^{|\boldsymbol{\mu}_i|} p_{\mu, z, s_i, \mu_{i, j}}
        \right),
    \end{multline*}
where $p_{\mu, z, i}$ is the probability that state $i$ is the initial state, $q_{\mu, z, i, j}$ is the transition probability from state $i$ to state $j$, and $p_{\mu, z, i, j}$ is the probability of medication $j$ conditioned on states $i$ and $z$.

\subsection{Estimation}

The estimation of the model parameters follows the standard Expectation Maximization (EM) procedure (see e.g. \cite{Moon1996-kb}).
The hyperparameters of the model are the number of top-level latent states $|Z|$ and the number of HMM states for the \textsf{Medications} sequence, $\mathcal{C}_S$.
In order to choose these 2 values, we compute the Bayesian Information Criterion (BIC), which penalizes the model fit by a function of the number of parameters: $BIC\left(d\right) = d\ln N - 2\ln f\left(\boldsymbol{y}\right)$, where $d$ is the total parameter count and $N$ is the number of episodes.
We perform a 2D grid search over $|Z|$ and $\mathcal{C}_S$ and compute the BIC to select these hyperparameters.

\subsection{Inference} \label{sec:inference}

%Using a trained model, we can perform an inference procedure to compute conditional likelihoods.
%Given a partial sequence of the episode up until a given time this procedure computes the likelihoods of possible continuations of the episode.
%This can be used, for example, to infer the final lengths of the sequences, or to find the most likely future values of any sequence.

Given a trained model and input data, we can derive inference algorithms to compute the likelihood of an event of interest.
This can include estimation of the final sequence length (e.g., length of \textsf{Beds} or \textsf{Laboratory Tests}) or prediction of specific future observations.

Of particular interest, is the inference of future events given a partial set of EHR data.
The collection of data for an episode up until time $t$ is
    $$
        \boldsymbol{y}_{[1\ldots t]} = 
        \left(\phi_a, \phi_s, \phi_d, \boldsymbol{\beta}_{[1\ldots t]}, \boldsymbol{\alpha}, \boldsymbol{\lambda}_{[1\ldots t]}, \boldsymbol{\nu}_{[1\ldots t]}, \boldsymbol{\mu}_{[1\ldots t]} \right),
    $$
where $\boldsymbol{x}_{[1\ldots t]} = \{ \boldsymbol{x}_1, \ldots, \boldsymbol{x}_w \}$ for the largest $w$ such that $t_{x, w} < t$ for component $x$.
We do not include \textsf{Discharge Diagnoses} or \textsf{Death} as inputs, since they are unknown until the end of the episode.

The central quantity that is needed to compute is
%inference problem using this model is to compute 
the conditional distribution of the latent variable $Z$ given the partial data: $f\left( Z=z | \boldsymbol{y}_{[1\ldots t]} \right)$.
%Once this distribution is determined, it is used in
%can be used to infer future values of the episode.
This distribution is the partial sequence likelihood weighted by cumulative Poisson probabilities,
    $$
        f\left( Z=z | \boldsymbol{y}_{[1\ldots t]} \right) = 
        p_z
        f\left(\boldsymbol{y}_{[1\ldots t]} | Z=z\right)
        \prod_x e^{-l_{x, z}}\sum_{k=t}^\infty \frac{l_{x, z}^k}{k!}.
    $$
%See XXX for the derivation of this expression.
In this expression, we evaluate the likelihood of the partial data for each $z$ and multiply by the remaining cumulative probability of future sequence lengths.
%This accounts for the likelihood of future values in each of the sequences.
This accounts for the lengths of the input sequences that have different Poisson parameters for each latent state,
and can be approximated easily as the values in the Poisson distribution approach zero.
%For example, assume we have trained a model in which the \textsf{Beds} sequence has a mean length of $l_{\beta,z}=4$ for some component $Z=z$.
%Then, given a partial \textsf{Beds} sequence with a length of 1, this cumulative value will be $\approx 0.91$.
%On the other hand, if the input is a partial \textsf{Beds} sequence with length 6, this value becomes $\approx 0.11$, indicating that this longer sequence is less likely to occur for this $z$.
%This calculation aggregates information from all of the episode data elements $x$ and their lengths to infer $Z$.

Once this distribution over $Z$ is computed using the partial input sequence $\boldsymbol{y}_{[1\ldots t]}$, we can infer quantities of interest.
%This is accomplished by integrating out, or summing, over all possible future values, $f\left( Z=z | \boldsymbol{y}_{[1\ldots t]} \right) = \sum_{\boldsymbol{y}_{[t, \ldots]}} f\left( Z=z | \boldsymbol{y}_{[1\ldots t]}, \boldsymbol{y}_{[t\ldots]} \right)$, where $\boldsymbol{y}_{[t,\ldots]}$ represents a collection of sequences in the episode starting after time $t$.
To infer the length of sequence $x$ we compute the probability of the sequence length given the partial sequence data,
    %$$
    %    f\left(k|\boldsymbol{x}_{[1\ldots t]}; \boldsymbol{l}_{\cdot} \right) = 
    %    \sum_z 
    %    f\left( Z=z | \boldsymbol{x}_{[1\ldots t]} \right)
    %    \frac{l_{\cdot, z}^k}{k!}
    %    e^{-l_{\cdot, z}},
    %$$
    $$
        f\left(k_x|\boldsymbol{y}_{[1\ldots t]}\right) = 
        \sum_z 
        f\left( Z=z | \boldsymbol{y}_{[1\ldots t]} \right)
        \frac{l_{x, z}^k}{k!}
        e^{-l_{x, z}}.
    $$
%where $\boldsymbol{l}_{\cdot}$ is the vector of Poisson rate parameters for component $\cdot$.
%See XXX for a derivation of this expression.
%Note that all data elements contribute to the calculation of $f\left( Z=z | \boldsymbol{y}_{[1\ldots t]} \right)$ and therefore the length distribution of this one sequence.
%For example, to compute the likelihood that the \textsf{Beds} sequence length is 6, the calculation of $f\left(Z=z|\boldsymbol{y}_{[1...t]}\right)$ uses the \textsf{Age}, \textsf{Sex}, \textsf{Admission Diagnoses}, \textsf{Laboratory Tests}, \textsf{Neurological Tests}, \textsf{Beds} and \textsf{Medications} up until time $t$.

To infer future values, the latent state probabilities are used as coefficients to mix over component distributions.
In general, inferring the distribution of component $\boldsymbol{x}$ can be accomplished by,
    $$
        f\left(\boldsymbol{x}|\boldsymbol{y}_{[1\ldots t]}\right) = 
        \sum_z
        f\left( Z=z | \boldsymbol{y}_{[1\ldots t]} \right)
        f\left(\boldsymbol{x} | Z=z\right).
    $$
%See XXX for the derivation of this expression.
%For example, the probability of \textsf{Death} is,
%    $$
%        f\left(\phi_d = 1\right) = 
%        \sum_z
%        f\left( Z=z | \boldsymbol{y}_{[1\ldots t]} %\right)
%        \left(1-p_{d,z}\right).
%    $$

\subsubsection{Individualized Sequence Length Prediction}

Individualized point predictions for sequence lengths may be computed by finding the mode of the resulting distributions,
    $$
        \hat{k}_{x,t} = \arg \max_k f\left(k|\boldsymbol{y}_{[1\ldots t]} \right).
    $$

\subsubsection{Population Level Inference} \label{sec:pred}

%As described above, this model is capable of inferring future quantities for a single episode.
%In this section, we provide inference expressions for populations as a cohort.
%This approach can be used to address the question of resource allocation for a group of subjects, rather than combining individual subject-level predictions.
%[Should we compare population to individualized in our results?]
%In the case of sequence length inference, we accomplish this by computing the sum of the expected value for each subject.
Considering a sequence $\boldsymbol{x}$ and using data up until time $t$, the expected value of the length of this sequence for subject $i$ is
    $$
        E_{x, i, t} = \sum_{k=0}^\infty k f\left(k|\boldsymbol{y}_{i, [1\ldots t]}\right).
    $$
%The summation terms will decrease to 0 as $k$ increases since we are evaluating a mixture of Poisson distributions for each $k$.
%Therefore, in practice, we can compute terms of this summation until the contributions become negligible.
The total population level sequence length estimate is computed by summing across episodes,
    $$
        E_{x, t} = \sum_i E_{x, i, t}.
    $$
This is the expected value of the total length across all subjects.
%[FIX may need to formalize this].

We are also interested in inferring the presence of a specific item in a sequence.
In this work, we will infer the presence of future ICU in the \textsf{Bed} sequence.
In order to do this, we first infer the \textsf{Bed} sequence distribution using episode data up until time $t$,
%[FIXME this gives us the $p$ and $q$ parameters, need to index by episode]
    $$
        f\left(\boldsymbol{\beta}\right) = 
        \sum_z
        f\left( Z=z | \boldsymbol{y}_{[1\ldots t]} \right)
        f\left(\boldsymbol{\beta} | Z=z\right).
    $$
%[FIX the rest of this subsection needs to be clarified and improved].
Under this model, the likelihood that ICU will not occur from time $t$ to time $t + s$ is 
%[FIXME need to index by episode]
    $$
        \sum_z \frac{l_{\beta, z}^{t + s}}{\left(t + s\right)!}
        \prod_{t'=1}^s q_{t + t', z},
    $$
where $q_{t, z}$ is the probability that the sequence at time $t$ 
%[FIXME sequence element, not time] 
is not ICU given $Z=z$, 
%[FIXME need to index by episode]
    $$
        q_{t,z} = 
        \frac{l_{\beta, z}^t}{t!}e^{-l_{\beta, z}}
        p_{\beta,z, \beta_{t-1}} 
        \sum_{i \neq ICU}
        q_{\beta, z, \beta_{t-1}, i}.
    $$
Under all future times, we can compute, 
%[FIXME need to index by episode]
    $$
        g = \sum_z \sum_{s=1}^\infty \frac{l_{\beta, z}^{t + s}}{\left(t + s\right)!}
        \prod_{t'=1}^s q_{t + t', z},
    $$
which is the probability that ICU will not occur in the future.
Our probability of interest is the ICU will occur, which is $1-g$.
This can be viewed as a Bernoulli trial in which the probability of success (ICU) changes at every step.
The population inference is computed by summing this probability over the episodes.

%\section{Model Training and Episode Profiling} %\label{sec:training}

%In this section, we describe the results of model training and explore the resulting model.

%    \begin{figure}[ht]
%        \centering
%        \begin{subfigure}[b]{0.45\textwidth}
%            \includegraphics[width=\textwidth]{figs/model_bic.png}
%            \caption{BIC values for all trained models}
%        \end{subfigure}
%        \begin{subfigure}[b]{0.45\textwidth}
%            \includegraphics[width=\textwidth]{figs/model_bic_zoom.png}
%            \caption{Zooming in on the local minima}
%        \end{subfigure}
%        \caption{Bayesian Information Criterion (BIC) computed for a series of models with varying hyperparameters. The hyperparameters control the number of latent mixture components and the number of latent states in the Medication sequences.}
%        \label{fig:modelsel}
%    \end{figure}

%[Exploring the model]

\section{Results}

From our dataset of 244,248 episodes, we randomly selected 80\% for training.
The two hyperparameters, $|Z|$ and $\mathcal{C}_S$ were selected by performing a linear search and computing the BIC on the training set as described in Section \ref{sec:methods}.
%Figure \ref{fig:modelsel} shows the 
BIC values were computed on a grid consisting of $|Z|$ taking values 1, 10, 20, 30, 40, 50, 60, 70, 80, 90, 100; and $\mathcal{C}_S$ taking values 
%1, 2, 5, 10, 15.
1, 5, 10, 15.
This resulted in 55 trained models.
%From the Figure, we can see that t
The model order $|Z|=30$ and $\mathcal{C}_S = 5$ attained the minimum BIC value over these models.
%The $\mathcal{C}_S$ values of 5, 10, and 15 are similar for this value of $|Z|$.
%To err on the side of parsimony
%We selected the values: $|Z|=30, \mathcal{C}_S = 5$.

%In this Section we present inference results for sequence length at the individual and population levels, and ICU presence.

\subsection{Individualized Inference of Sequence Length}

%We used the trained model selected in Section \ref{sec:training} and the inference approach described in Section \ref{sec:methods} to infer the length of the \textsf{Beds}, \textsf{Laboratory Tests}, and \textsf{Medications} sequences at every timepoint in the test set.
%This set of timepoints consists of all times when new data arrive in any of the sequences.
%We use this maximal set of timepoints for an episode which contains all of the timepoints at which new data arrive.
%This set is different for different episodes.

Estimated sequence lengths are computed at every available timepoint for each test episode.
Our method is compared to two baseline approaches.
The first, ``Constant", uses the training population average sequence lengths, which is 2.68 for \textsf{Beds}, 197 for \textsf{Laboratory Tests}, and 15.2 for \textsf{Medications} per episode.
The second, ``Hold Last", outputs the current sequence length at every timepoint (i.e., estimates that the final sequence length is equal to the present sequence length).

Given the true value for the length of component $x$, $k_x^\star$, we quantify the Absolute Error (AE),
    $$
        e_{x,t} = |\hat{k}_{x,t} - k_x^\star|,
    $$
where $\hat{k}_{x,t}$ is the prediction at time $t$ and $k_x^\star$ is the final true value.
%When formulating average performance over the test cohort, we can average the AE either over timepoints, $\frac{1}{N|\mathcal{T}|} \sum_n \sum_{t \in \mathcal{T}_n} e_{x,t}^{(n)}$, or episodes, $\frac{1}{N} \sum_n \frac{1}{|\mathcal{T}_n|} \sum_{t \in \mathcal{T}_n} e_{x,t}^{(n)}$, where $e_{x,t}^{(n)}$ is the result for episode $n$ and $\mathcal{T}_n$ is set of timepoints for episode $n$.

The results in Table \ref{tbl:indiv_seq} show the AE per episode for the three methods.
Values in parenthesis are the percentage improvement over the ``Constant"  method performance.
%[Also need per timepoint results].

%\begin{table}[ht]
%    \centering
%     \begin{tabular}{|c c c|} 
%     \hline
%     Method & AE per timepoint & AE per episode \\ [0.5ex] 
%     \hline
%        Constant (2.68) & 1.68 (0) & 0.93 (0) \\
%        Hold Last  & 1.03 (37) & 0.72 (23) \\
%        Model  & 0.85 (49) & 0.45 (52) \\
%     \hline
%     \end{tabular}
%     \caption{Number of Beds Prediction Results. Values is parenthesis are percentage improvement over the Constant baseline approach.}
%     \label{tbl:results-bed}
%\end{table}

\begin{table}[ht]
    \centering
     \begin{tabular}{|c c c c|} 
     \hline
     Method & \textsf{Beds} & \textsf{Laboratory Tests} & \textsf{Medications} \\ [0.5ex] 
     \hline
        Constant & 0.93 (0)     & 129 (0)   & 6.9 (0) \\
        Hold Last  & 0.72 (23)  & 85 (34)   & 6.2 (10) \\
        Model  & 0.45 (52)      & 73 (43)   & 4.3 (38) \\
     \hline
     \end{tabular}
     \caption{Absolute error of sequence length prediction results. Values are the average absolute error per episode and values in parenthesis are percentage improvement over the Constant baseline approach.}
     \label{tbl:indiv_seq}
\end{table}

\subsection{Population-Level Inference}

%Here we presents population level inference results for inference of sequence length and ICU presence in the \textsf{Beds} sequence.
%As opposed to the individual-level inference in the preceding section, for population-level inference we do not make decisions for each episode and therefore need a new performance metric.
For population-level inference we compute the average absolute error, 
    $$
        e = 100\frac{|y_{pred}\left(t\right) - y\left(t\right)|}{\sum_t v\left(t\right)},
    $$
where $y_{pred}\left(t\right)$ is the prediction at time $t$, $y\left(t\right)$ is the true value at time $t$, and $v\left(t\right)$ is the number of episodes available at time $t$.
The values $y_{pred}$, $y\left(t\right)$, and $v\left(t\right)$ are accumulated over all episodes in the test set.
%For example, for ICU prediction, at time $t$ we have predicted that their will be a total of $y_{pred}\left(t\right)$ of the episodes with ICU in their \textsf{Bed} sequence.
%At this time $t$, there are $v\left(t\right)$ active episodes, and of these $y\left(t\right)$ of them will contain ICU in their \textsf{Bed} sequences.

\subsubsection{Sequence Length Inference}

For this problem, we compute the population level inference for sequence length as given in Section \ref{sec:methods}.
We compare our method with the baseline population rate for each sequence, as determined from the training data.
Table \ref{tbl:popseq} shows the results.

\begin{table}[ht] 
    \centering
     \begin{tabular}{|c c c c|} 
     \hline
     Method & \textsf{Beds} & \textsf{Laboratory Tests} & \textsf{Medications} \\ [0.5ex] 
     \hline
        Baseline & 3.27 & 364.06 & 22.33 \\
        Model  & 0.64 & 214.84 & 8.43 \\
     \hline
     \end{tabular}
     \caption{Average absolute error for the baseline and model approaches in inferring the length of each sequence.}
     \label{tbl:popseq}
\end{table}

\subsubsection{ICU Presence Inference}

For ICU inference, we compare our approach to the population baseline rate of ICU admittance in the training set.
For reference, we also compute the error for the two edge cases of strictly predicting that ICU will and will not exist.
We call these methods ``All" and ``None", respectively.
%Figure \ref{fig:res_icu} shows the population totals over time on a logarithmic scale.
%From this Figure, the 'Total' curve shows the total number of episodes.
%This is decreasing since the episodes have variable lengths.
%The 'ICU' curve is the true number of episodes that will have ICU in their \textsf{Bed} sequences.
%%Note that this curve merges with the 'Total' curve, since all of the very long episodes will contain ICU.
%The 'Baseline' curve shows a fixed proportion from the 'Total' curve, and the 'Model' curve shows our results.
%    \begin{figure}[ht]
%    \centering
%    \includegraphics[width=0.5\textwidth]{figs/pop_icu.png}
%     \caption{Logarithmic time scale showing the total number of episodes, true number of ICU, the baseline and model results.}
%     \label{fig:res_icu}
%    \end{figure}
%Summarizing these results,
Table \ref{tbl:res_icu} shows the averaged absolute error as defined above in this Section.
    
    \begin{table}[ht] 
    \centering
     \begin{tabular}{|c c|} 
     \hline
     Method & Average AE \\ [0.5ex] 
     \hline
        All     & 60.86 \\
        None    & 39.14 \\
        Baseline & 15.40 \\
        Model   & 12.32 \\
     \hline
     \end{tabular}
     \caption{Average absolute error for ICU inference.}
     \label{tbl:res_icu}
\end{table}

\section{Discussion and Conclusions}

In this work, we defined a model for episodic EHR data containing mixed sequences and static information.
The model is a mixture over probability distributions tailored to each data type, including collections and sequences.

Expressions were derived that enable inference of future values given partial input sequence data.
These are based on inferring the underlying latent variable of the mixture model.
Inference depends not only on the values, but also on the lengths of the input sequences.

We trained the model on data from KPNC.
Sequence length prediction and presence of future ICU in the \textsf{Beds} sequence was performed.
We find that our approach outperforms the population average baseline, indicating that the model is capturing individualized information and is capable of generalizing beyond the training set.

In Table \ref{tbl:indiv_seq} we see that the model outperforms the baseline methods for all three of the sequences.
The Hold Last approach produces lower AE than the Constant approach.
This is because there is substantial spread in the sequence lengths across individuals and the population average does not provide an accurate estimate.
These results show that our method is leveraging individualized information contained within the sequences to predict sequence length.

Table \ref{tbl:popseq} shows that at the population level, our model also outperforms the baseline.
In this approach, we are estimating the total sequence length across all subjects the test set.
The sequence length of the \textsf{Laboratory Tests} were more difficult to predict than the \textsf{Beds} or \textsf{Medications}.
Since these sequences are very long, it may be that the dynamics are more difficult to learn compared to other sequences.
Prediction of sequence length may be a useful task for resource planning.
Being able to predict utilization of resources using individualized information rather than population averages can lead to more accurate estimates of short-term future resource needs.

Prediction of future ICU presence (Table \ref{tbl:res_icu}) shows that our model produces lower error than the baseline methods.
The Baseline method is significantly better than the edge cases of assuming all patients will be in the ICU (All) and that no patients will be in the ICU (None).
This indicates that the average rate of ICU presence may be a relatively stable value.
There was an decrease in error, however, when using our model, showing that the model is capturing patterns that are informative for predicting ICU presence.
This problem could also be a test case of potential use in resource planning, where more specific information may be needed in addition to the total sequence length.

Although the inference algorithms are computationally fast to compute, the training is expensive.
Being able to update the model given new training data may be an important feature to develop for practical adoption of this method.
Various techniques exist for online training of latent variable models, and the exploration of those methods for this problem is left as future work.

% Acknowledgements should go at the end, before appendices and references

%\section*{Acknowledgment}


\bibliographystyle{IEEEtran}
\bibliography{refs}

\end{document}